\documentclass[aps,prl,preprint,superscriptaddress,floatfix]{revtex4-1}
\usepackage{amsmath,amsfonts,amssymb}
\usepackage{graphicx,color}
\usepackage{epstopdf}
\usepackage{bm}

\begin{document}
%
%
%
\title{
Coherence Resonance and Stochastic Resonance in an Excitable Semiconductor Superlattice }
\author{Emanuel Mompo}
\affiliation{Gregorio Mill\'an Institute for Fluid Dynamics, Nanoscience
and Industrial Mathematics, and Department of Materials Science and Engineering and Chemical Engineering, Universidad Carlos III de Madrid, 28911 Legan\'{e}s, Spain}

\author{Miguel Ruiz-Garcia}
\affiliation{Gregorio Mill\'an Institute for Fluid Dynamics, Nanoscience
and Industrial Mathematics, and Department of Materials Science and Engineering and Chemical Engineering, Universidad Carlos III de Madrid, 28911 Legan\'{e}s, Spain}

\author{Manuel Carretero}
\affiliation{Gregorio Mill\'an Institute for Fluid Dynamics, Nanoscience
and Industrial Mathematics, and Department of Materials Science and Engineering and Chemical Engineering, Universidad Carlos III de Madrid, 28911 Legan\'{e}s, Spain}

\author{Holger T. Grahn}
\affiliation{Paul-Drude-Institut f\"ur Festk\"orperelektronik, Leibniz-Institut im
Forschungsverbund Berlin e. V., Hausvogteiplatz 5--7, 10117 Berlin, Germany}

\author{Yaohui Zhang}
\affiliation{ Key Laboratory of Nanodevices and Applications, Suzhou Institute of Nano-Tech and Nano-Bionics, Chinese Academy of Sciences, Suzhou 215123, China}
\affiliation{Department of Physics, University of Science and Technology of Defense, Changsha, 413000, China}

\author{Luis L. Bonilla}\email{bonilla@ing.uc3m.es}
\affiliation{Gregorio Mill\'an Institute for Fluid Dynamics, Nanoscience
and Industrial Mathematics, and Department of Materials Science and Engineering and Chemical Engineering, Universidad Carlos III de Madrid, 28911 Legan\'{e}s, Spain}
%
%
%
\begin{abstract}
Collective electron transport causes a weakly coupled semiconductor superlattice under dc voltage bias to be an excitable system with $2N+2$ degrees of freedom: electron densities and fields at $N$ superlattice periods plus the total current and the field at the injector. External noise of sufficient amplitude induces regular current self-oscillations (coherence resonance) in states that are stationary in the absence of noise. Numerical simulations show that these oscillations are due to the repeated nucleation and motion of charge dipole waves that form at the emitter when the current falls below a critical value. At the critical current, the well-to-well tunneling current intersects the contact load line. We have determined the device-dependent critical current for the coherence resonance from experiments and numerical simulations. We have also described through numerical simulations how a coherence resonance triggers a stochastic resonance when its oscillation mode becomes locked to a weak ac external voltage signal. Our results agree with the experimental observations. 
\end{abstract}
%
\maketitle

Constructive effects of noise include superresolution in time reversal acoustics \cite{der95,blo02}, signal enhancement due to stochastic resonance (SR) \cite{mcn88,gam98,bla05,bur08}, coherence resonance (CR) \cite{hu93,pik97,lee05,hiz06}, etc. In nonlinear excitable systems \cite{kee98}, noise of appropriate strength can trigger coherent oscillations (CR) and enhance the signal-to-noise ratio of a periodically driven bistable system (SR). These constructive effects of noise are typically demonstrated in few-degrees-of-freedom systems amenable to analytical and simple numerical studies, e.g., a particle in a double-well potential under white noise and ac driving forces in the SR case \cite{gam98} and an excitable system described by the FitzHugh-Nagumo equation in the CR case \cite{pik97,lee05}.

Technologically relevant devices are often complex and harder to characterize, yet they may exhibit CRs as well as SRs. A case in point are excitable semiconductor superlattices (SSLs). Because of sequential tunneling electron transport (STET), voltage-biased, doped, weakly coupled SSLs can be modeled as nonlinear systems with many degrees of freedom. They exhibit excitable or oscillatory behavior depending on the driving and configuration parameters \cite{bon02,BGr05}. For large doping densities, SSLs have multistable stationary states that produce sawtooth-like current-voltage characteristics under dc voltage bias. A square voltage pulse may induce excitability \cite{luo98} visualized by a large current spike caused by the formation at the cathode and motion towards the anode of a charge dipole wave (CDW) \cite{BGr05,ama01}. After the wave disappears, there remains a stable static state consisting of a lower electric-field domain near the cathode followed by a higher field domain that extends to the anode. For lower doping densities, the SSL total current (TC) may oscillate periodically in time due to repeated CDW formation and motion \cite{BGr05}. Depending on the cathode conductivity, doping density, and temperature, voltage intervals of stable stationary states may be followed by intervals of stable oscillatory states \cite{hiz06}. Based on numerical simulations, a CR has been predicted \cite{hiz06} and observed in experiments on GaAs/AlAs SL at low temperatures \cite{hua14}.

Recently, under dc voltage bias, spontaneously chaotic \cite{hua12,li13}, periodic, and quasiperiodic \cite{hua13} self-sustained current oscillations have been observed in GaAs/Al$_{0.45}$Ga$_{0.55}$As 50-period SLs at room temperature. Noise may induce or enhance chaotic oscillations over a wider voltage bias range provided its amplitude is sufficiently large and its bandwidth is much smaller than the oscillation frequency \cite{yin17}. Numerical simulations show that thermal and shot noise enhance deterministic spontaneous chaos in a STET model of a SSL of 50 identical spatial periods \cite{alv14,bon17}. For shorter SSLs, theory predicts enhanced deterministic chaos due to a Feigenbaum period-doubling cascade in certain voltage intervals \cite{rui17}. Reference~\cite{ess18} studies how variations in basic design parameters influence chaos.

In this Letter, we study the CR in dc voltage-biased SSLs at room temperature driven by external noise having a bandwith larger than the oscillation frequency. We also study the SR when a small ac voltage is added. The corresponding experimental results are presented in Ref.~\cite{sha18}. For a dc voltage bias, when the current drops below a critical value, the external noise may produce large current spikes due to the formation of a CDW at the injector that propagates toward the collector and disappears there. The value of the critical TC (CTC) to trigger a CDW is given by the intersection of the well-to-well sequential current density with the injector load line (current density versus local electric field). Both these device-specific functions cannot be directly determined from experiments. However, the CTC can be extracted from numerical simulations of the theoretical model by studying the ratio of the standard deviation to the mean duration of large current spikes. As the noise amplitude increases, a coherent oscillation develops, which corresponds to a minimum of the standard deviation of interspike time intervals divided by the mean interspike time. Similar to this numerically demonstrated CR, noise may enhance a weak ac signal, thereby demonstrating a SR. Experiments confirm these predictions \cite{sha18}.

\begin{figure}
\includegraphics[width=8cm]{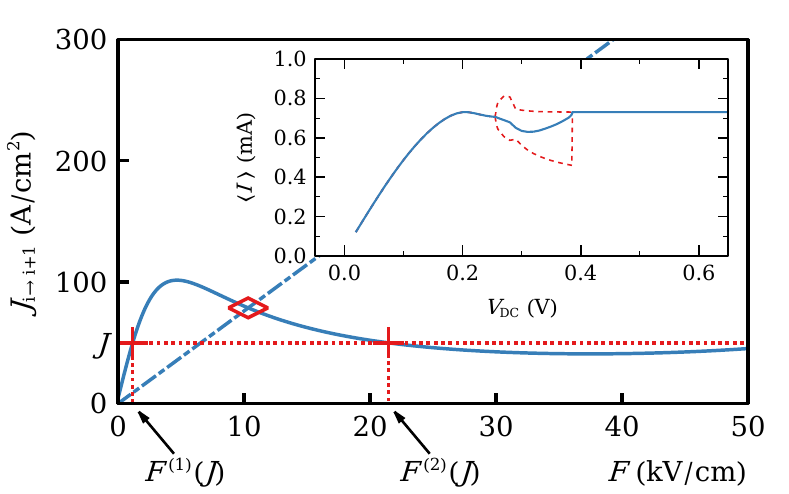}
\caption{
Current-field characteristics (solid line) and injector load line (dot-dashed line) for $F_i=F$ and $n_i=N_D$. The first intersection point yields the CTC (rhombus), $J_{\rm cr}=78.7959$ A/cm$^2$ and field, $F_{\rm cr}=10.3265$ kV/cm. Inset: $I$--$V$ characteristics indicating maximum and minimum values of the current in the oscillatory regime (dotted lines). }
\label{fig1}
\end{figure}

\noindent {\em Model.} The electric field $-F_i$ and the two-dimensional electron density $n_i$ at well $i$ $(i=1,\ldots,N$) satisfy \cite{bon02,BGr05,bon17,bon94,bon00,suppl}
\begin{eqnarray}
\epsilon \frac{dF_i}{dt}+J_{i\rightarrow i+1}=J(t),\label{eq1}\\
n_i=N_D +\frac{\epsilon}{e}(F_i-F_{i-1}),\label{eq2}
\end{eqnarray}
where $-e<0$, $\epsilon$, $N_D$, $J(t)$ and $J_{i\rightarrow i+1}$ are the electron charge, SSL average permittivity, doping density, TC density, and tunneling current density from well $i$ to $i+1$, respectively \cite{bon02,BGr05,bon17}:
\begin{eqnarray}
J_{i\rightarrow i+1}=\frac{e n_i}{l} v^{(f)}(F_i)-J_{i\rightarrow i+1}^{-}(F_i,n_{i+1},T),\label{eq3}\\
J_{i\rightarrow i+1}^{-}(F_i,n_{i+1},T)=\frac{em^* k_B T}{\pi \hbar^2 l}v^{(f)}(F_i)\nonumber\\
\times\ln \left[1+ e^{-\frac{eF_i l}{k_B T}}\left( e^{\frac{\pi \hbar^2 n_{i+1}}{m^* k_B T}} -1\right)  \right]\!.\label{eq4}
\end{eqnarray}
The function $v^{(f)}(F_i)$ has peaks corresponding to the discrete energy levels in every well \cite{bon02,BGr05,bon17} (45, 173 and 346 meV, for a 7 nm GaAs/4 nm Al$_{0.45}$Ga$_{0.55}$As SL \cite{hua12,li13,yin17,sha18}). The barrier height is 388 meV. The mesa cross section is a square with a 30 $\mu$m side length and $N=50$ \cite{sha18}. $m^*$, $l$, $k_B$, $T$ are the average effective mass, SSL period, Boltzmann constant, and lattice temperature, respectively \cite{bon02}. Voltage bias and boundary conditions are \cite{kro68,bon00,bon02,BGr05,bon17,yin17}
\begin{eqnarray}
l\sum_{i=1}^{N}F_i= V+\eta(t), \quad \eta(t)=\eta_\text{th}(t)+\eta_{\text{c}}(t),\label{eq5}
\end{eqnarray}
$J_{0 \rightarrow 1}=\sigma_0 F_0$ and $J_{N \rightarrow N+1} =\sigma_0 \frac{n_N}{N_D}F_N$. Figure \ref{fig1} shows the tunneling current density versus the constant field $F_i=F$ for $n_i=N_D$ and also the injector load line (dot-dashed line). In Eq.~\eqref{eq5}, the voltage $V$ may comprise the dc voltage bias $V_\text{dc}$ and an ac signal $V_\text{ac}=V_\text{sin}\sin (2\pi\nu t)$. The voltage noise $\eta(t)$ has two components: (i) $\eta_\text{th}(t)$, which is related to the noise of the source, and (ii) the external noise $\eta_{\text{c}}(t)$. $\eta_\text{th}(t)$ is simulated by picking random numbers every $5$$\times$$10^{-11}$~s from a zero mean distribution with a standard deviation of $2$ mV  \cite{yin17}. $\eta_{c}(t)$ is a white noise with bandwidth of 1 GHz and tunable amplitude $V_\text{noise}^\text{rms}$. These noise values have been selected so that the results of the numerical simulations of the model agree qualitatively with the results of the experiments reported in Refs.~\cite{yin17,sha18}. We have ignored the smaller value of $\eta_\text{th}(t)$ at the SSL wells \cite{alv14,bon17}.

\begin{figure}
\hspace{-1.25cm} \includegraphics[width=8cm]{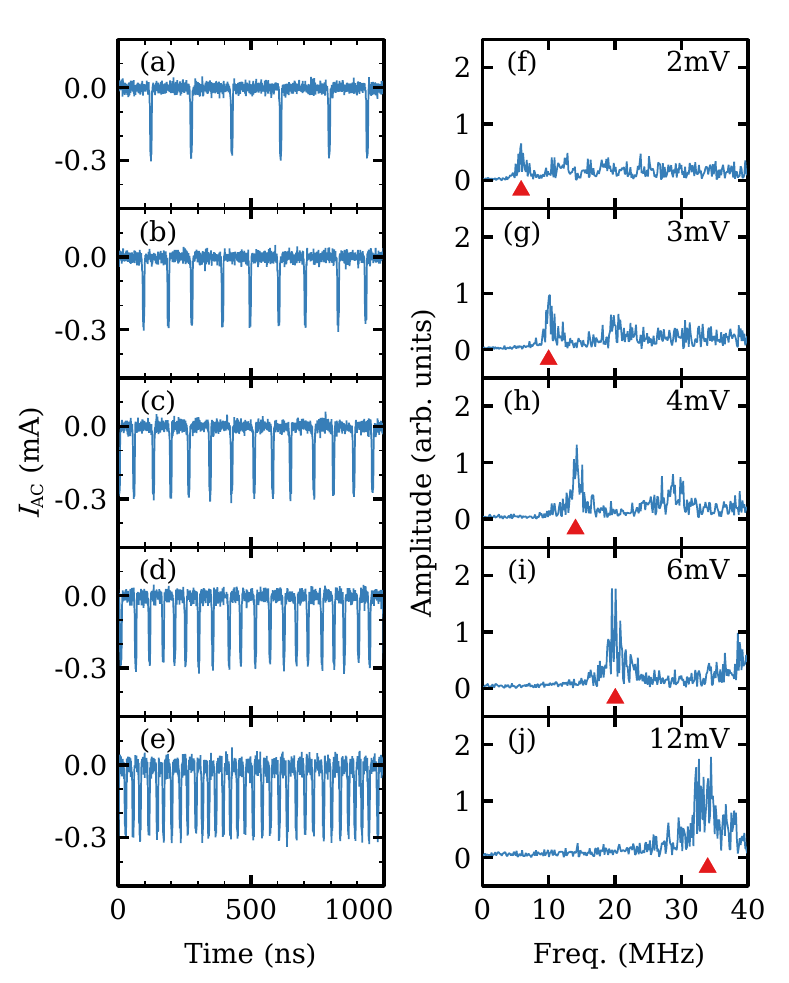}
\caption{Coherence resonance: (a)--(e) ac components of the TC versus time and (f)--(j) corresponding frequency spectra (interspike average frequency marked by a triangle) for different noise amplitudes at $V_\text{dc}=0.387$ V. Values of  $V_\text{noise}^\text{rms}$ are 2, 3, 4, 6, and 12 mV. Current traces have been shifted to have zero current at the stationary state. }
\label{fig2}
\end{figure}

\noindent {\em Results.} Eqs.~\eqref{eq1}--\eqref{eq5} yield predictions that are in {\em qualitative agreement} with experiments. Typically the TC and the frequency of TC self-oscillations (TCSO) are lower than observed \cite{BGr05}, which we shall consider when comparing with experiments. For deterministic dc voltage bias, TCSO begin as a supercritical Hopf bifurcation at $V_\text{dc}=0.255$ V and end at a saddle-node, infinite-period bifurcation (SNIPER) at $V_\text{dc}=0.385$ V (cf.\ inset of Fig.~\ref{fig1}). The experiments exhibit the same scenario \cite{sha18}.
\begin{figure}
\includegraphics[width=8cm]{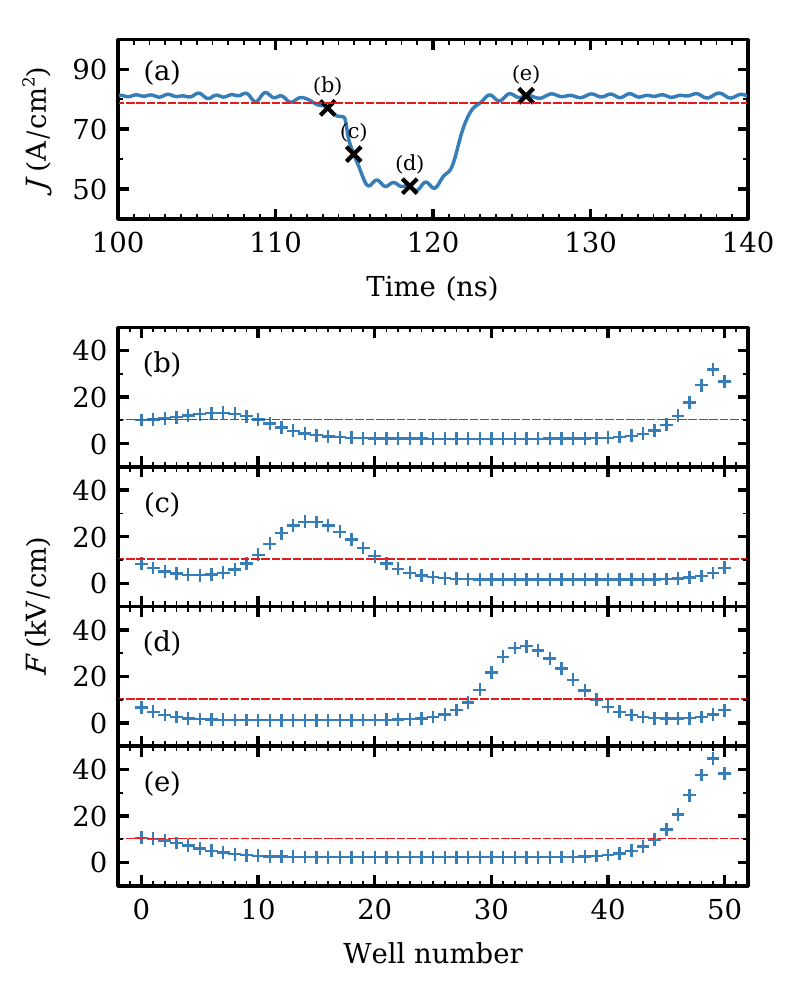}
\caption{(a) TC density versus time for $V_\text{dc}=0.387$ V, $\eta_\text{th}=0$, and $V_\text{noise}^\text{rms}=3$ mV. (b)--(e) Field profiles at the times marked in (a). Dashed lines indicate the critical current and field. See also movie in the Supplement Material \cite{suppl}.}
\label{fig3}
\end{figure}

Adding noise with increasing amplitude at $V_\text{dc}=0.387$ V, TCSO appear as shown in Figs.~\ref{fig2}(a)--\ref{fig2}(e) for $V_\text{noise}^\text{rms}>1.4$ mV indicating the presence of a CR. The CR frequency follows the interspike average frequency [marked by triangles in Figs.~\ref{fig2}(f)--\ref{fig2}(j)], which increases with increasing $V_\text{noise}^\text{rms}$. For $V_\text{noise}^\text{rms}< 2$ mV, the TC presents a rapid small-amplitude oscillation (caused by the noise) and large spikes separated by long-time intervals. Between spikes, the TC is close to a constant value slightly above $J_{\rm cr}$ defined in Fig.~\ref{fig1}. Figure \ref{fig3}(a) shows an enlarged view of $J(t)$ for an interval containing one large current spike with field profiles shown in Figs.~\ref{fig3}(b)--\ref{fig3}(e). Outside the spike, the corresponding field profile is quasistationary [cf. Fig.~\ref{fig3}(e)]:  $F_i\approx F_\text{cr}$ near the injector, then $F_i$ decreases to $F^{(1)}(J)$, stays there for several periods, and increases again near the collector. As shown in Figs.~\ref{fig3}(b)--\ref{fig3}(d), each large current spike corresponds to CDW creation and motion when $J$ decreases and stays below $J_{\rm cr}$ for some time. As $J<J_{\rm cr}$, the high-field region near the collector tries to move out leaving a field $F^{(1)}(J)$ on its wake. However,  the total area under the electric field profile is conserved on average according to Eq. \eqref{eq5}. As the pulse near the collector departs, the lost area has to be compensated by launching a new CDW at the injector, which causes the TC to decrease as Figs.~\ref{fig3}(a)--\ref{fig3}(c) show. When the CDW arrives at the collector and starts disappearing, $J$ increases up to its stationary value (except for  noise-produced small-amplitude oscillations), and $F_i$ becomes quasistationary [cf. Fig.~\ref{fig3}(e)].

\begin{figure}
 \includegraphics[width=8cm]{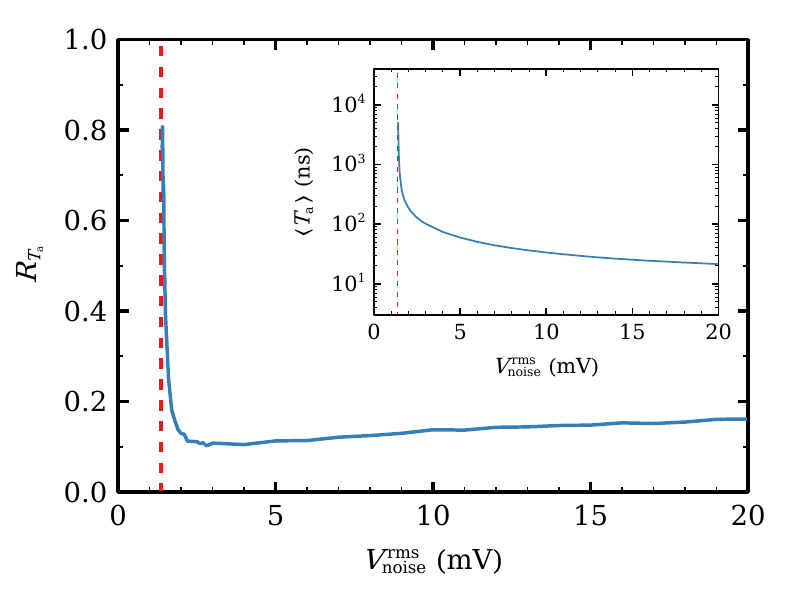}
\caption{Normalized standard deviation $R_{T_a}$ versus $V_\text{noise}^\text{rms}$ ($\eta_\text{th}=0$). Inset: mean interspike interval $\langle T_a\rangle$ versus $V_\text{noise}^\text{rms}$. The vertical asymptotes (dashed lines) occur at $V_\text{noise}^\text{rms}=1.365$ mV.}
\label{fig4}
\end{figure}

Figure \ref{fig4} depicts the normalized standard deviation $R_{T_a}=\sqrt{\langle T_a^2\rangle-\langle T_a\rangle^2}/\langle T_a\rangle$ of the interspike time interval, $T_a$, versus $V_\text{noise}^\text{rms}$. It exhibits a minimum after an abrupt drop followed by a smooth increase. This behavior is expected for voltages close to a SNIPER bifurcation \cite{hiz06}. The mean interspike interval $\langle T_a\rangle$ shown in the inset of Fig.~\ref{fig4} first decreases from infinity at $V_\text{noise}^\text{rms}=1.365$ mV and rather more smoothly for $V_\text{noise}^\text{rms}>2$ mV. This behavior is typical of a CR and agrees qualitatively with the experimental results \cite{sha18}.

\begin{figure}
 \includegraphics[width=8cm]{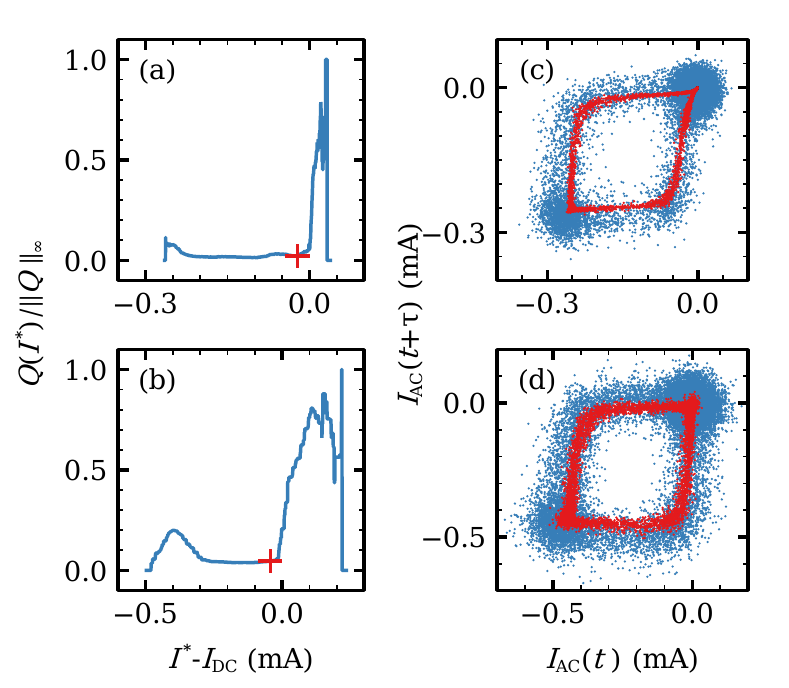}
\caption{Normalized standard deviation to mean ratio of large current spike duration, $Q(I^*)$, versus $I^*-I_\text{dc}$ for a CR from: (a) numerical simulations with $V_\text{dc}=0.387$ V, $\eta_\text{c}=6$ mV, and $\eta_\text{th}=0$; (b) experimental data with $V_\text{dc}=0.773$ V and $V_\text{noise}^\text{rms}=8.288$ mV. The crosses in (a) and (b) mark $I_\text{cr}$. Standard deviations and means are taken over the union set of all disjoint time intervals lasting more than $t^*$ (one third of the average duration of a large current spike) during which the current is $I(t)<I^*$. Simulations and experimental data yield $t^*=10$ and 3 ns, respectively. Attractor of the CR in the $(I_\text{ac}(t),I_\text{ac}(t+\tau))$ phase plane of embedded coordinates \cite{strogatz}: (c) numerical simulations with $V_\text{dc}=0.387$ V, $\tau=1.642$ ns, $\eta_\text{c}=8$ mV, $\eta_\text{th}=2$ mV; (d) experimental data with $\tau=0.884$ ns and other parameters as in (b). The sharper red attractors in (c) and (d) are obtained by noise filtering with a 8-level Haar wavelet \cite{mal08}.}\label{fig5}
\end{figure}

We now estimate the CTC for triggering new pulses by comparing experimental with simulated results. This device-dependent quantity (cf. Fig. \ref{fig1}) has great importance for theory. We select data at voltages slightly larger than the SNIPER bifurcation, 0.387 V (theory) and 0.773 V (experiments) \cite{sha18}. The larger experimental value is due to  voltage drops at the contact region and at the 50 Ohm impedance of the oscilloscope. The stationary state TC is obtained as the average current between two large spikes when the external noise is so low that few large spikes exist. For the model, we know the CTC exact value, 0.709 mA, which is 97\% of the stationary state TC. From the experiments, 97\% of the stationary state TC (1.448~mA) is 1.404~mA, which is our estimated CTC. This value is confirmed by comparing the theoretical and experimental normalized ratio of the standard deviation of the duration of large long-lasting current peaks to their mean duration. Figure \ref{fig5} shows this comparison. The normalized ratio versus current, CTC location, and even the shape of the CR attractor are qualitatively quite similar (cf. Ref.~\cite{suppl}).

\begin{figure}
 \includegraphics[width=8cm]{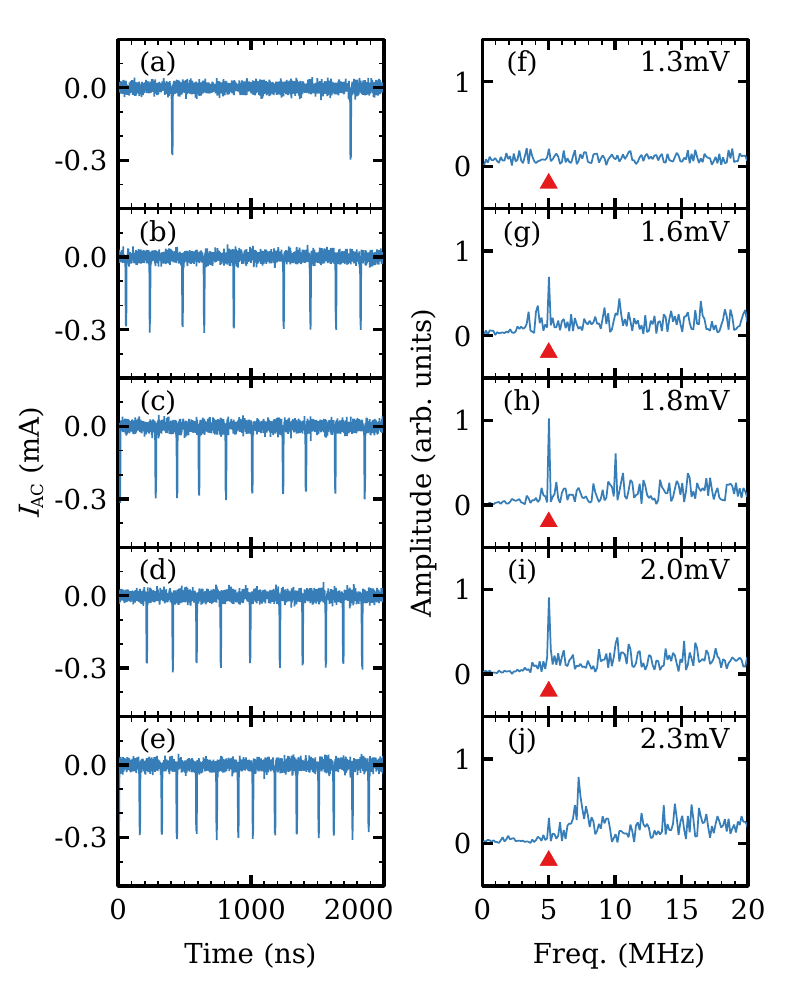}
\caption{Stochastic resonance: (a)--(e) ac components of the SSL current versus time and (f)--(j) corresponding frequency spectra (frequency of ac signal marked by a triangle) for different noise amplitudes at $V_\text{dc}=0.387$ V and a sinusoidal ac signal of frequency $\nu=$ 5 MHz and $V_\text{sin}=0.646$ mV. The values of  $V_\text{noise}^\text{rms}$ are 1.4, 1.7, 1.9, 2.1, and 2.3 mV. }
\label{fig6}
\end{figure}

\begin{figure}
 \includegraphics[width=8cm]{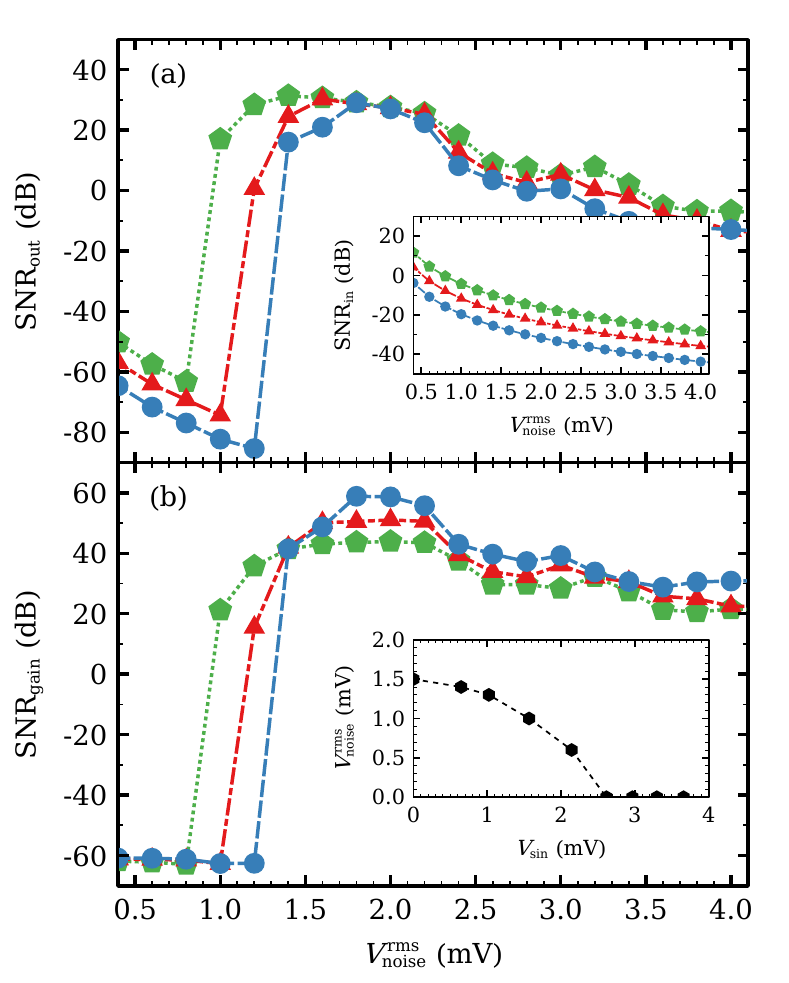}
\caption{(a) SNR$_\text{out}$ (inset: SNR$_\text{in}$) and (b) SNR$_\text{gain}$ versus $V_\text{noise}^\text{rms}$ (varied from 0.5 to 4 mV) for $V_\text{sin}=0.646$ (circles), 1.022 (triangles) and 1.736 mV (pentagons). The inset of (b) shows the values of $V_\text{noise}^\text{rms}$ needed to trigger periodic TCSO versus $V_\text{sin}$.}
\label{fig7}
\end{figure}

Figures \ref{fig6}(a)--\ref{fig6}(e) show the result of adding a small amplitude ac signal to the dc voltage with a frequency within the CR range and then increasing the noise amplitude. Isolated current spikes separated irregularly by long intervals appear for $V_\text{noise}^\text{rms}<1.4$ mV. With increasing noise amplitudes, the SSL oscillates at a frequency locked with that of the ac signal, as shown in Figs.~\ref{fig6}(g)--\ref{fig6}(i). At larger $V_\text{noise}^\text{rms}$, the main frequency increases and ceases to be locked to that of the ac signal, as shown in Fig.~\ref{fig6}(j). This is the signature of a SR. Figures \ref{fig7}(a) and \ref{fig7}(b) show the output signal-to-noise ratio SNR$_\text{out}$ and the gain of the signal-to-noise ratio SNR$_\text{gain}\!=$ SNR$_\text{out}/$SNR$_\text{in}$, respectively, of the SSL under SR \cite{suppl}. Experiments confirm that this SR exists \cite{sha18}. However, the enhancement of the signal-to-noise ratio (more than 100 dB) is larger in the simulations than that observed in experiments (more than 30 dB) \cite{sha18}, not surprisingly as our idealized model does not include many noise sources of the actual experimental setup. As shown in the inset of Fig.~\ref{fig7}(b), the necessary noise amplitude for frequency locking is smaller than that needed for a CR when $V_\text{sin}=0$. $V_\text{noise}^\text{rms}$ decreases with $V_\text{sin}$, as observed in the experiments \cite{sha18}.

In conclusion, by numerical simulations of a STET model of SSLs at room temperature, we have found a CR and a SR. The CR is due to repeated and coherent generation of CDWs at the injector when the amplitude of external noise surpasses a certain threshold. For the first time, we have estimated the value of the critical current necessary to trigger waves from experimental data. When we add an external ac signal with a frequency within that of the CR, the SSL phase locks to the ac signal, even if the latter is weak \cite{suppl}. Our simulations agree qualitatively with the experimental results \cite{sha18} and confirm that SSLs under SR can act as lock-in amplifiers.

The authors thank the Ministerio de Econom\'\i a y Competitividad of Spain (Grants No. MTM2014-56948-C2-2-P and No. MTM2017-84446-C2-2-R), the Strategic Leading Science and Technology Special Program of the Chinese Academy of
Sciences (Grant No. XDA06010705), the National Natural Science Foundation of China (Grants No. 61070040 and No. 61204093), the National Key Research and Development Program of China (Grant No. 2016YFE0129400), and the Exploration Project (Grant No. 7131266) for financial support. E.M. acknowledges support from the Ministerio de Econom\'\i a y Competitividad of Spain through the 2015 Formaci\'on de Doctores program cofinanced by the European Social Fund. M.R.-G. acknowledges support from Ministerio de Educaci\'on, Cultura y Deporte of Spain through the Formaci\'on de Profesorado Universitario program.

\appendix
\section{Supplemental material}

\subsection{Model}
Let us revise the premises of the model we use for charge transport in semiconductor superlattices (SSLs). For weakly coupled SSLs, the theory is essentially at the same stage as described in our review \cite{BGr05} and updated by considering internal noise in Ref.~\cite{bon17}. The SSL miniband widths are small compared to the broadening of the energy levels due to scattering and to the typical values of the electrostatic energy per SL period, $eFl$ ($-e<0$ is the electron charge, $-F$ electric field, and $l$ the SSL period). Then the escape time from a quantum well is much larger than the scattering time, which implies that the electron distribution in the wells is in local equilibrium \cite{bon94}. The dielectric relaxation time, in which the current density across the SSL reacts to sudden changes in the electric field profile, is typically larger than the escape time. Therefore, we can assume that the tunneling current density between quantum wells is stationary on the longer time scale of the dielectric relaxation time \cite{BGr05,bon17}. A minimal theory of charge transport in weakly coupled SSLs should therefore specify (i) which slowly varying magnitudes characterize the local equilibrium distribution function in the wells (at least the electric field and the electrochemical potential or the electron density), (ii) the equations relating these magnitudes (e.g. the charge continuity and the Poisson equation), and (iii) how to close these equations by calculating the necessary relations between magnitudes (e.g. the stationary tunneling current between adjacent wells). This means that the space variables in the charge continuity and Poisson equations are discrete (each superlattice period is represented by an index $i$, $i=1,\ldots,N$) and that the tunneling current appearing in the charge continuity equation is calculated using a number of approximations (see \cite{BGr05,bon17}). The resulting equations are 
\begin{eqnarray}
\epsilon \frac{dF_i}{dt}+J_{i\rightarrow i+1}=J(t),\label{eqs1}\\
n_i=N_D +\frac{\epsilon}{e}(F_i-F_{i-1}),\label{eqs2}\\
J_{i\rightarrow i+1}=\frac{e n_i}{l} v^{(f)}(F_i)-J_{i\rightarrow i+1}^{-}(F_i,n_{i+1},T),\label{eqs3}
\end{eqnarray}
\begin{eqnarray}J_{i\rightarrow i+1}^{-}(F_i,n_{i+1},T)=\frac{em^* k_B T}{\pi \hbar^2 l}v^{(f)}(F_i)\nonumber\\
\times\ln \left[1+ e^{-\frac{eF_i l}{k_B T}}\left( e^{\frac{\pi \hbar^2 n_{i+1}}{m^* k_B T}} -1\right)  \right]\!.\label{eqs4}
\end{eqnarray}
Here $m^*$, $\epsilon$, $n_i$, $N_D$, $J(t)$ and $J_{i\rightarrow i+1}$, $k_B$, and $T$ are the SSL average effective mass and permittivity, the two-dimensional (2D) electron density, the 2D average doping density, the total current density and the tunneling current density from well $i$ to $i+1$, the Boltzmann constant, and the lattice temperature, respectively \cite{BGr05,bon17}. The forward electron velocity $v^{(f)}(F_i)$ is a function with peaks corresponding to the discrete energy levels $\epsilon_{C_j}$, $j=1,\ldots,n$ ($n=3$ for the SSL configuration we use), in every well:
\begin{eqnarray}
&&v^{(f)} (F_i) = \sum_{j=1}^n \frac{\frac{\hbar^3 l (\gamma_{C1} + \gamma_{C_j})}{2m^{*2}}\,
{\cal T}_i (\epsilon_{C1}) }{ (\epsilon_{C1} - \epsilon_{C_j} + eF_il)^2 + (\gamma_{C1} + \gamma_{C_j})^2},
\label{eqs5}\\
&&{\cal T}_i (\epsilon) \!=\! \frac{16 k_i^2 k_{i+1}^2 \alpha_i^2
 (k_i^2 + \alpha_i^2)^{-1} (k_{i+1}^2 + \alpha_i^2)^{-1} }
{(d_W + \alpha_{i-1}^{-1} + \alpha_i^{-1}) (d_W + \alpha_{i+1}^{-1} + \alpha_i^{-1})
 e^{2 \alpha_i d_B} },\, \label{eqs6}\\
&&\hbar k_i =  \sqrt{2 m^* \epsilon}, \quad \hbar k_{i+1} = \sqrt{2 m^* (\epsilon + e l F_i)}, \label{eqs7}\\
&&\hbar \alpha_{i-1} = \sqrt{2 m^* \left[ e V_B + e \left( d_B + \frac{d_W}{2} \right) F_i
- \epsilon\right]}, \label{eqs8}\\
&&\hbar \alpha_i =  \sqrt{2 m^* \left[ e V_B- \frac{e d_W F_i }{2} - \epsilon\right]}, \label{eqs9}\\
&&\hbar \alpha_{i+1} = \sqrt{2 m^* \left[ e V_B - e \left( d_B + \frac{3d_W}{2} \right) F_i
- \epsilon\right]}. \label{eqs10}
\end{eqnarray}
Here $d_B$, $d_W$, with $l=d_B+d_W$, are the barrier and well widths, respectively, and $eV_B$ is  the barrier height. The energy broadening parameters of the energy levels are $\gamma_{C_j}$. The energy levels are calculated by solving a Kronig-Penney model for the particular SSL configuration under study. The values of the parameters we use in our numerical simulations are listed in Table \ref{t1}. 

\begin{table}[ht]
\begin{center}\begin{tabular}{ccccccccccccc}
 \hline
$d_W$ &$d_B$ &$N_D$& $V_B$&$\epsilon_{C_1}$&$\epsilon_{C_2}$&$\epsilon_{C_3}$&$\gamma_{C_1}$&$\gamma_{C_2}$&$\gamma_{C_3}$ &$\epsilon$ & $m^*$&$\sigma_0$\\
nm  & nm  &$\frac{10^{10}}{\text{cm}^2}$ &  $mV$ &  meV& meV& meV& meV& meV& meV&$\epsilon_0$ & $m_\text{e}$&$\frac{\text{A}}{\text{Vm}}$\\ 
7 & 4 &6&  388&  45& 173&346& 2.5&8&24&12.1 &0.1 &0.763\\
 \hline
\end{tabular}
\end{center}
\caption{Parameters used to solve the model equations with $N=50$ at $T=295$ K. The SSL cross section is a square of 30 $\mu$m side and $m_e$ is the electron mass.}
\label{t1}
\end{table}

To solve Eqs.~\eqref{eqs1}-\eqref{eqs2}, we need additional boundary and bias conditions. Under voltage bias conditions, we have
\begin{eqnarray}
l\sum_{i=1}^{N}F_i= V+\eta(t), \quad \eta(t)=\eta_{th}(t)+\eta_{c}(t). \label{eqs11}
\end{eqnarray}
Here $V$ is the voltage between the SSL ends and $\eta(t)$ is a voltage noise described in the main text. Equation \eqref{eqs1} should include an internal noise in the tunneling current coming from shot and thermal noise \cite{bon17}. However, we ignore this noise because it is much smaller than the controllable external noise $\eta_c(t)$ and the internal noises due to circuit elements in the experimental setup, $\eta_{th}(t)$.  This is more realistic than the way external noise was included in Ref.~\cite{hiz06}: they had zero noise in Eq.~\eqref{eqs11} and included tunable external white noise within each SSL well. Their simulations showed a coherence resonance as the amplitude of their external noise increased. 

To solve Eqs.~\eqref{eqs1} and \eqref{eqs2}, we need the fields at the contact regions, $F_0$ and $F_N$, respectively. We lack detailed modeling of the contact regions, but we can propose relations between the current density and the field there, in the same spirit as Kroemer's imperfect boundary conditions for the Gunn effect \cite{kro68}. Some time ago, we derived boundary conditions at the contacts by using the transfer Hamiltonian formalism \cite{bon00}. The tunneling currents through the barriers separating the contact regions from the SSL are:
\begin{equation}
J_{0 \rightarrow 1}=j(F_0), \quad J_{N \rightarrow N+1} =n_{N}\, w(F_N),\label{eqs12}
\end{equation}
where $j(F_0)$ and $w(F_N)$ are some functions specified in Ref.~\cite{bon00}. The relevant features of these theoretically based boundary conditions are the value of the critical current density and critical field at which the contact load line $j(F_0)$ intersects the tunneling current density Eq.~\eqref{eqs3} in appropriate units \cite{bon00}. At the critical current, the injecting contact emits a charge dipole wave into the superlattice. This intersection point is the same as produced by Ohm's law, $j(F_0)=\sigma_0 F_0$, with appropriately chosen contact conductivity $\sigma_0$. The boundary condition at the collector does not influence significantly the SSL dynamics, and a linear function $w(F_N)=\sigma_0 F_N/N_D$ is often used \cite{BGr05}. We can only infer indirectly that the boundary conditions are reasonable by comparing qualitative predictions of the theory with the experimental results. Our numerical simulations of the model with linear $j(F_0)$ and $w(F_N)$ show that these simplified boundary conditions are sufficient to qualitatively explain the experimental findings. 

\subsection{Contact conductivity versus voltage phase diagram}
 
 \begin{figure}
 \includegraphics[width=8cm]{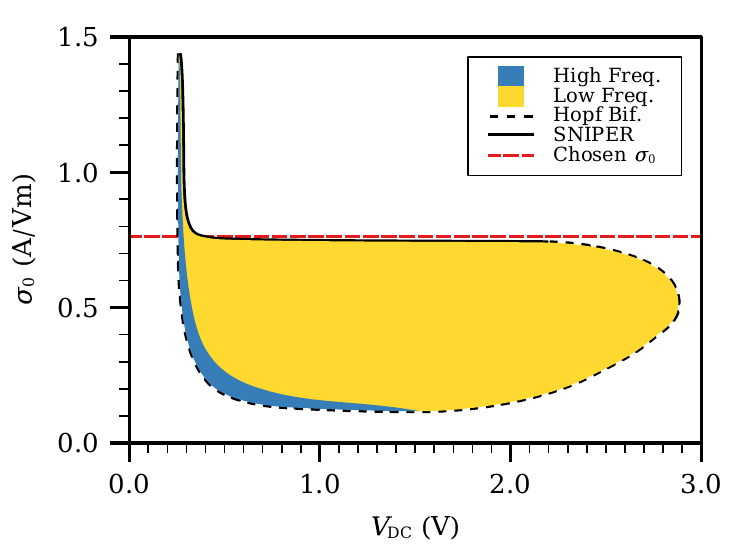}
\caption{Phase diagram of injector contact conductivity versus dc voltage exhibiting a bounded region of current self-oscillations. At the dashed boundary line, the self-oscillations appear as Hopf bifurcations from the stationary field profile which is linearly stable outside the bounded region. The continuous boundary line corresponds to saddle-node infinite period bifurcations. The contact conductivity used in the simulations is marked as a horizontal line.}
\label{figsup1}
\end{figure}
\begin{figure}
 \includegraphics[width=8cm]{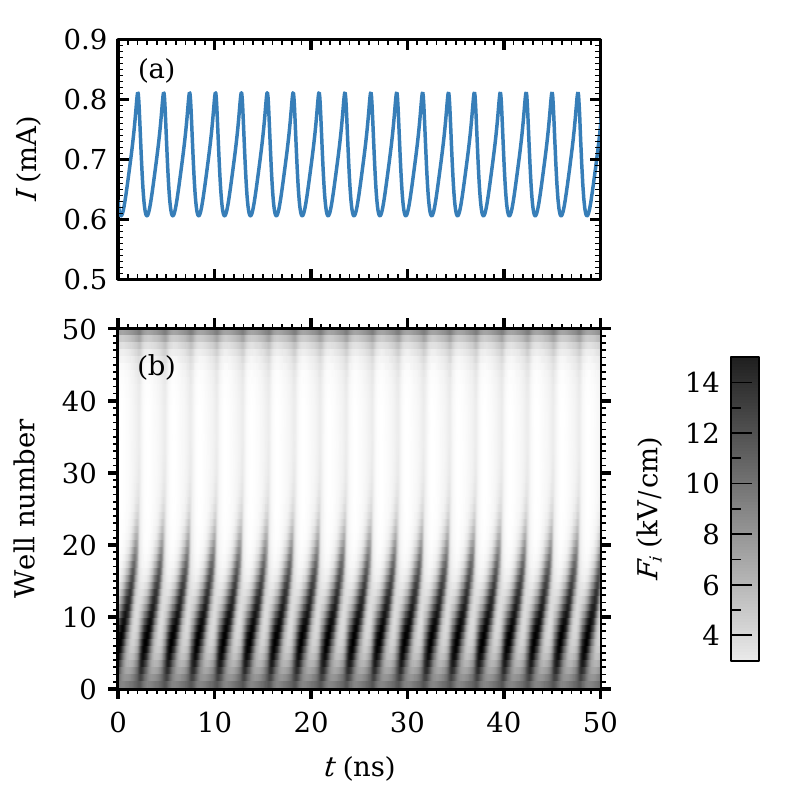}
\caption{(a) Total current versus time of a high-frequency self-oscillation. (b) Evolution of the field density plot for the same times as in (a). Bias $V=V_\text{dc}=0.27$ V (no noise).}
\label{figsup2}
\end{figure}
\begin{figure}
\includegraphics[width=8cm]{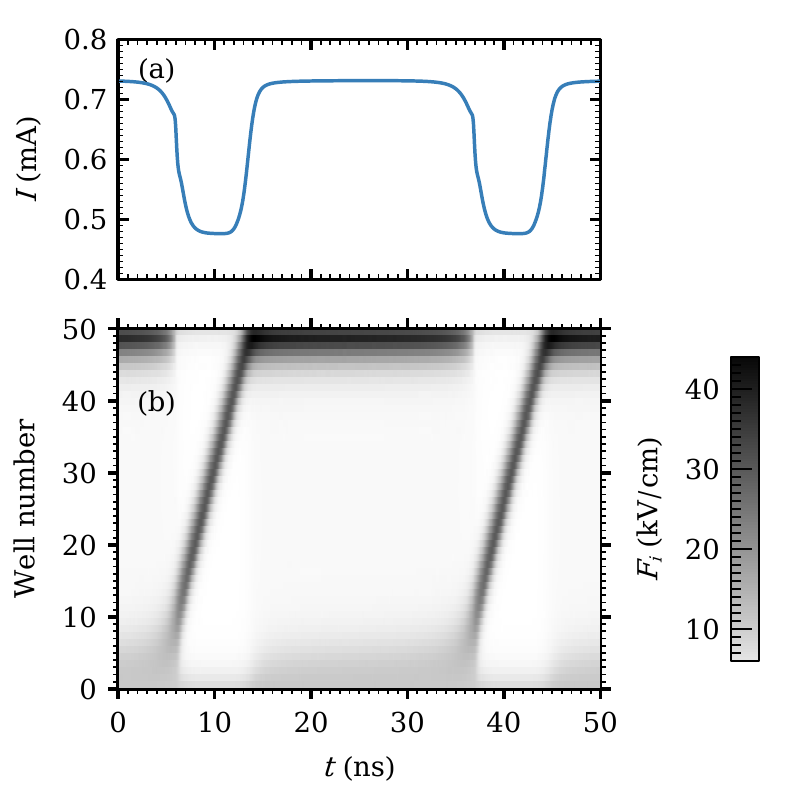}
\caption{(a) Total current versus time of a low-frequency self-oscillation. (b) Evolution of the field density plot for the same times as in (a). Bias $V=V_\text{dc}=0.36$ V (no noise).}
\label{figsup3}
\end{figure}
Let us consider a pure dc voltage bias in the absence of noise: $\eta=0$ in Eq.~\eqref{eqs11}. The stationary field profile is linearly stable for values of the contact conductivity and dc voltage outside some bounded region depicted in Fig.~\ref{figsup1}. As the parameters cross the boundary line, self-oscillations of the total current appear. Our numerical simulations show that the self-oscillations appear as Hopf bifurcations across the dashed part of the boundary in Fig.~\ref{figsup1} and as saddle-node infinite period (SNIPER) bifurcations across the continuous line. Thus, the conductivity of the injecting contact, $\sigma_0$, selects the type of bifurcation. Note that this bifurcation scenario (Hopf followed by SNIPER bifurcation as the dc voltage increases) differs from that in Fig.~3 of Ref.~\cite{hiz06}, where the SNIPER is at the onset of the interval of self-oscillations, not at its end.

In addition, for a given contact conductivity, the amplitude and frequency of the current self-oscillations change with voltage. For dc voltage next to the Hopf bifurcation, the self-oscillations are due to creation and recycling of charge dipole waves over a finite region of the SSL that is close to the injecting contact (cf. Fig.~\ref{figsup2}). The frequency of these self-oscillations is larger than that of self-oscillations due to creation, motion and recycling of charge dipole waves through the entire superlattice (cf. Fig.~\ref{figsup3}). The maxima of the total current are also larger for the high-frequency oscillations, as it can be observed in the inset of Fig.~1 of the main text and by comparing Figs.~\ref{figsup2}(a) and \ref{figsup3}(b). The low-frequency oscillations occur between a certain critical voltage and the SNIPER bifurcation voltage, and they are marked in Fig.~\ref{figsup1}. The critical voltage for dipole waves to move across the entire SSL, $V_\text{d}$, can also be observed  in the inset of Fig.~1 of the main text as a dip in the maxima of the total current. As shown in Fig.~\ref{figsup3}, the corresponding current versus time plots are different for the low-frequency oscillations: the current stays for a long time at a certain value (the field profile is almost stationary), then rapidly drops, and then goes up again (the dipole wave disappears at the collector, a new wave is created at the injector, it moves to the collector and is stuck there reproducing the quasistationary state). As the voltage approaches the SNIPER point, this long time between current drops goes to infinity (thus the frequency is low), but the amplitude stays the same. Note that, {\em for the chosen contact conductivity, the voltage interval of low-frequency self-oscillations, $V_\text{SNIPER}-V_d$, is much wider than that of high-frequency self-oscillations, $V_d-V_\text{Hopf}$.}

What is the effect of adding a small voltage noise when $V_\text{dc}$ is fixed outside the interval of current self-oscillations but near the SNIPER point? Adding a small noise can make the voltage $V$ affecting the device to go back to the aforementioned voltage interval where current self-oscillations still occur. Then the current spikes triggered by the noise are no different than the deterministic ones. Thus, the smoothness of the mean interspike interval, $\langle T_a\rangle$, can be explained by the relative length of $\upsilon=(V_\text{SNIPER}-V_d)/(V_\text{SNIPER}-V_\text{Hopf})$. The larger (smaller) $\upsilon$ is, the smoother (abrupter) the changes for $\langle T_a\rangle$ will be. According to Supplementary Fig.~\ref{figsup1}, $\upsilon$ can be made smaller by increasing the contact conductivity $\sigma_0$. However, then $(V_\text{SNIPER}-V_\text{Hopf})$ becomes too small, which is why we did not select a larger value of $\sigma_0$.

\subsection{Estimation of the critical current}
The critical current at which a charge dipole wave is emitted by the injecting contact is given by the intersection of the well-to-well tunneling current density (for constant field and electron density equal to the doping density) with the contact load line. It is marked by a rhombus in Fig.~1 of the main text. Numerical simulations produce the accompanying movie {\em current$_-$spike$_-$and$_-$dipole.mp4}, which illustrates the creation of a wave when there is a pronounced spike of the current and the current spends sufficient time below its critical value. Figure 3 of the main text contains four snapshots of this movie. 

Here we want to estimate the critical current directly from the simulations and then use the same criteria to determine the critical current from the experimental data. We know that the critical current is 97\% of that corresponding to the stationary state. We want to extract this value from the simulations of the model including noise at $V_\text{dc}=0.387$ V by noticing that the large long-lasting current spikes occur when the total current spends sufficient time below $I_\text{cr}$. How do we characterize this current from statistics of the large current spikes? 

Firstly, we note that the average duration of a large current spike in Fig.~2 of the main text is 10 ns. For a given time trace, let $t^*$ be one third of this average time and let $\mathcal{A}(I^*)$ be the union set of all disjoint time intervals $\mathcal{I}$ (lasting more than $t^*$) during which the current is $I(t)<I^*$:
\begin{widetext}
\begin{equation}
\mathcal{A}(I^*) = \{
I_i\subset \mathbb{R}: I(t)<I^*\ \forall t\in I_i,\ m(I_i) > t^*,\ I_i\cap I_j=\emptyset, \ \forall i\neq j
\}, \label{eq13}
\end{equation}
\end{widetext}
where $m(I_i)$ is the length (measure) of the interval $I_i$. Since we want to study the regularity of the intervals $I_i$, the observable will be their durations, i.e., $X=(m(I_1), m(I_2),\ldots)$. Then the ratio between the standard deviation of the time intervals $I_i$ to their mean is 
\begin{equation}
Q(I^*)=\frac{\text{std}(X)}{\text{mean}(X)}.\label{eqs14}
\end{equation}
Figure 5(a) of the main text depicts $Q(I^*)$ (normalized to have a maximum value of one) as a function of the current $I^*-I_\text{dc}$ as derived from numerical simulations. There is an almost flat region of $Q(I^*)$ between two hills. The leftmost hill corresponds to the small-amplitude noisy oscillations at the bottom of a large current spike as that of Fig.~3 of the main text. The rightmost hill corresponds to the small-amplitude noisy oscillations at the top of each large current spike. The rightmost region of large $Q(I^*)$ occurs for $I^*$ slightly larger than $I_\text{cr}$ (cf. Fig.~1 of the main text) marked by a cross in Fig.~5(a). This critical current is 97\% of $I_\text{dc}$. We use this fact to estimate $I_\text{cr}$ from the experiments.  

We now repeat the same construction using data obtained from the experiments, namely, from Fig.~2(a) of Ref.~\cite{sha18}. The average duration of a large current spike, 3 ns, which is shorter than in the simulations. Thus, $t^*$ is 1 ns for the experiments. As explained before, the model underestimates the frequency of the oscillations and overestimates the times involved in them. From the average value of the current during the small-amplitude oscillations between large current spikes, we can estimate the current at the stationary state, $I_\text{dc}$. The normalized ratio of the standard deviation to the mean duration of time intervals, $Q(I^*)$ given by Eq.~\eqref{eqs14}, is depicted in Fig.~5(b) of the main text. It has the same shape as $Q(I^*)$ in the simulations. The estimated critical current (97\% of $I_\text{dc}$) is also marked by a cross in Fig.~5(b). Note that this critical value is also slightly smaller than the beginning of the rightmost hill in the same figure. 

Further assurance that numerical simulations and data from experiments describe the same coherence resonance is obtained by visualizing the corresponding attractor. We use embedded coordinates and depict $(I_\text{ac}(t),I_\text{ac}(t+\tau))$ in Figs.~5(c) and 5(d) of the main text (appropriate values of $\tau$ for numerical simulations and experimental data are indicated in the caption). For both, theory and experiment, the shape of the attractor is roughly a square. From the numerical simulations, we know that the stationary state is in the upper right corner of the square, the traveling dipole wave in the lower left corner, and transitions between these states along the sides of the square. The wavelet filter used to refine the visualization data for the coherence resonance attractor is described in the Supplemental Ref.~\cite{mal08}.
 
 \subsection{Signal-to-noise ratio}
 Let $S_m$ be a measured signal that can be split as $S_m = S + \eta$, with $S$ the original signal and $\eta$ the noise, both known. Then the signal-to-noise ratio can be computed as $\text{SNR} = \text{rms}(S)^2/\text{rms}(\eta)^2$.

If there is no direct access to the decomposition (only $S_m$ is known), a noise filter $\mathfrak{F}$ can be used to obtain the following approximations: $S\approx \mathfrak{F}(S_m)$ and $\eta\approx S_m - \mathfrak{F}(S_m)$.

For $\text{SNR}_\text{in}$ in the main text, we use $S=V_\text{ac}=V_\text{sin}\sin(2\pi\nu t)$ and $\eta = \eta_\text{th}+\eta_\text{c}$. Note that $\text{rms}(V_\text{ac})=V_\text{sin}/\sqrt{2}$ and $\text{rms}(\eta)^2 = \text{rms}(\eta_\text{th})^2 + \text{rms}(\eta_\text{c})^2$, where $\text{rms}(\eta_\text{th})=2\text{mV}$ and $\text{rms}(\eta_\text{c})=V^\text{rms}_\text{noise}$.

For $\text{SNR}_\text{out}$, $S_m$ is the total current density $J$, and therefore we need a noise filter. For this, we use an 6-level Haar Wavelet \cite{mal08}.

In Fig.~7 of the main text, we use a logarithmic scale to depict the SNR, that is, we plot $20\log_{10}(\text{SNR})$.

 \subsection{Stochastic resonance}
\begin{figure}
 \includegraphics[width=8cm]{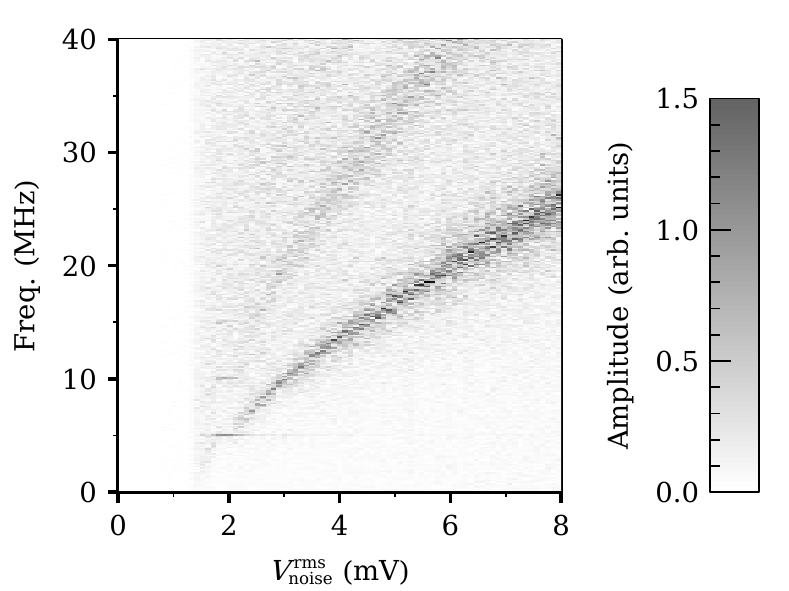}
\caption{Stochastic resonance: Density plot of the frequency spectra versus noise amplitude at $V_\text{dc}=0.387$ V and a sinusoidal ac signal of frequency 5 MHz and $V_\text{sin}=0.646$ mV.  These are the parameters used in Figure 6 of the main text.}
\label{figsup4}
\end{figure}

\begin{figure}
 \includegraphics[width=8cm]{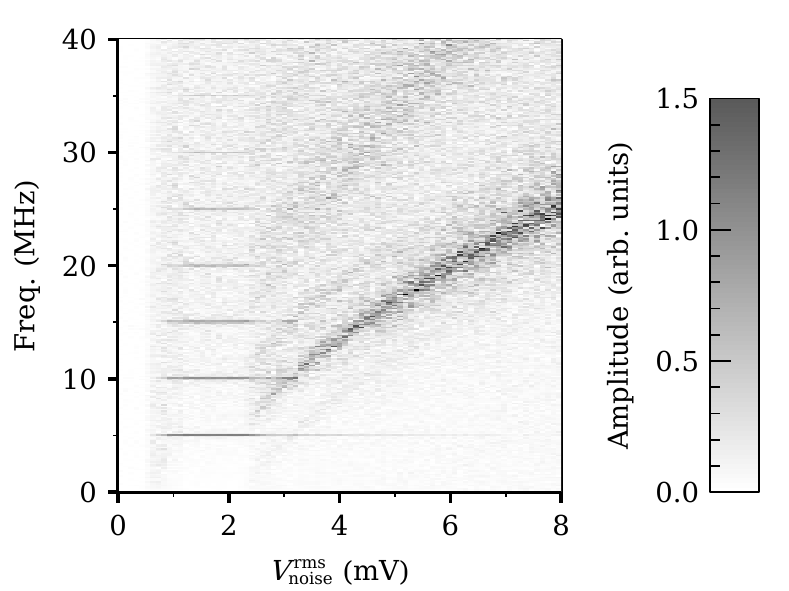}
\caption{Stochastic resonance: Density plot of the frequency spectra versus noise amplitude at $V_\text{dc}=0.387$ V and a sinusoidal ac signal of frequency 5 MHz and a larger value $V_\text{sin}=2.144$ mV. }
\label{figsup5}
\end{figure}
Supplemental Figs.~\ref{figsup4} and \ref{figsup5} show that the addition of an external ac signal with a frequency within that of the coherence resonance induces phase locking to the ac signal, even if the latter is weak, provided the noise amplitude is sufficiently large.

\subsection{Movie}
Movie {\em current$_-$spike$_-$and$_-$dipole.mp4} shows the appearance of a large current spike and the corresponding transition from the stationary state to a traveling dipole and back. Depicted are the total current density versus time (upper panel) and, synchronized with it, the field versus well number (lower panel). Dashed lines in the panels represent critical current density (upper panel) and critical field (lower panel). Parameters are as in Fig.~3 of the main text.

\end{document}